\documentclass[%
 reprint,
superscriptaddress,
nofootinbib,
longbibliography,
bibnotes,
 amsmath,amssymb,
 aps,
 pre,
]{revtex4-1}

\usepackage{graphicx}
\usepackage{dcolumn}
\usepackage{bm}
\usepackage{color}
\definecolor{OliveGreen}{rgb}{0,0.6,0}
\definecolor{auburn}{rgb}{0.43, 0.21, 0.1}
\definecolor{blue_violet}{rgb}{0.54, 0.17, 0.89}
\definecolor{hokie_maroon}{RGB}{99, 0, 49} 
\definecolor{hokie_orange}{RGB}{207, 69, 32} 

\usepackage{appendix}
\usepackage{chngcntr}
\usepackage{etoolbox}
\usepackage{lipsum}
\usepackage{mathtools}

\AtBeginEnvironment{appendices}{%
\addcontentsline{toc}{section}{Appendices}
\counterwithin{figure}{section}
\counterwithin{equation}{section}
\counterwithin{table}{section}
}

\begin{document}


\title{Origin of the Phase Separation into B2 and L2$_1$ Ordered Phases in the X--Al--Ti (X: Fe, Co, and Ni) Alloys from the First-principles Cluster Variation Method }

\author{Ryo Yamada}
\email{r-yamada@mat.eng.osaka-u.ac.jp}
\affiliation{Division of Materials and Manufacturing Science, Graduate School of Engineering, Osaka University, Suita, Osaka 565-0871, Japan}
\author{Tetsuo Mohri}
\email{tmohri@imr.tohoku.ac.jp}
\affiliation{Institute for Materials Research, Tohoku University, Sendai 980-8577, Japan}

\date{\today}

\begin{abstract}
The phase separation behaviors from the single B2 ordered phase into the two separate B2 and L2$_1$ ordered phases in the X--Al--Ti (X: Fe, Co, and Ni) alloys are analyzed using the cluster variation method (CVM) with the interaction energies evaluated from the electronic band structure calculations. The cubic approximation of the CVM is employed for the X$_2$Al$_{2-x}$Ti$_x$ ($0 \leq x \leq 2$) alloys limiting an interchange between Al and Ti atoms on the $\alpha$- and $\beta$-sublattices of the L2$_1$ ordered structure with the X atoms fixed on the $\gamma$-sublattice. The phase stabilities of the B2 and L2$_1$ structures are examined, and the phase diagrams at the pseudo-binary section, XAl--XTi, are determined. The two-phase regions of B2 and L2$_1$ phases, i.e., phase separation behavior, are successfully produced in both Co-- and Ni--Al--Ti alloy systems, whereas no phase separation is predicted in the Fe--Al--Ti alloy. The origin of the phase separation in the Co-- and Ni--Al--Ti alloys is, respectively, attributed to the mechanical instability and the combination of mechanical instability and chemical repulsions of unlike pairs. 
\end{abstract}

\pacs{Valid PACS appear here}
\maketitle

\section{\label{sec:level1}Introduction}
A decomposition of a solid solution into two separate phases in metallic alloy systems is one of the most well-known and -studied phenomena in materials science. In general, the decomposition is mechanically and/or chemically induced; when an initial solid solution becomes unstable or metastable in a given environment, such as high pressures and/or low temperatures, it decomposes into two different phases. This phenomenon is well-described by the empirical rules known as the Hume--Rothery's rules, where following three factors are used to evaluate a phase stability of a single solid solution in an alloy system \cite{haasen1978physical}: a difference of atomic radii, an electronegativity, and a valency electron concentration, each of which is estimated from constituent elements of the given alloy system. The former one and the latter two are, respectively, related to mechanically and chemically driven phase separations. 

The mechanically induced phase separation originates from mechanical instability. The mechanical instability can be readily detected from a pressure--volume ($P$--$V$) curve. The $P$--$V$ curve represents the second order derivative of the energy (or free energy), $E$, in terms of the volume, $V$, because $\partial^2 E/\partial V^2 = - \partial P/\partial V$. To be mechanically stable, a system has to satisfy the condition $\partial P/\partial V < 0$ at an equilibrium volume where the curve intersects with the horizontal axis, i.e., $P=P_{\mbox{\scriptsize{ext}}} \approx 0$. The concept of mechanical instability leading to the phase separation has been applied to a liquid--gas transformation \cite{girifalco2003statistical}, but is also applicable to a solid--solid transformation \cite{mohri2004first}.

The chemically driven phase separation, on the other hand, can be concisely described by the use of a 1st-nearest-neighbor pair interaction model. For example, when the two components of an A--B alloy have a stronger bonding for A--A and B--B pairs than for A--B (or B--A) pairs, the ground-state structure will be composed of two chemically distinct phases, A-rich and B-rich solid solutions. At relatively high temperatures, an entropic contribution becomes dominant, and a single solid solution (or a disordered phase) will get stable over the two separate phases. In other words, a high-temperature single solid solution decomposes into two different phases at low temperatures. 

In some alloy systems, a phase separation with an accompanying ordering has been observed (for examples, see Refs.\;\cite{ino1978pairwise,allen1976mechanisms,soffa1989decomposition}). This is known as a concurrent behavior of ordering and phase separation. The concurrent behavior in Fe--Be alloy, from a single solid solution into a solid solution and B2 ordered phase, has been successfully explained by considering further distant pair interactions \cite{ino1978pairwise}. In fact, it has been suggested that when the 1st-nearest-neighbor interactions prefer an ordering but the 2nd-neighbors promote a phase separation, there is a possibility of an occurrence of the concurrent behavior of ordering and phase separation \cite{inden1974ordering,ino1978pairwise}. 

In Co-- and Ni--Al--Ti ternary alloys, a similar phenomenon with the above mentioned concurrent behavior has been observed \cite{kainuma1997ordering,ishikawa2002ordering,ishikawa2002phase}. In these particular cases, a B2 ordered phase decomposes into B2 and L2$_1$ ordered phases. Interestingly, both B2 and L2$_1$ ordered phases have been observed in Fe--Al--Ti system as well, but the phase separation behavior has not \cite{kainuma1997ordering}. There are several works which suggest that the phase separation in the Ni--Al--Ti alloy is due to a lattice misfit between the B2 and L2$_1$ phases \cite{oh1997phase,enomoto1997analysis}; i.e., it is caused by the mechanical instability. 

\begin{figure}
\includegraphics[scale=0.38]{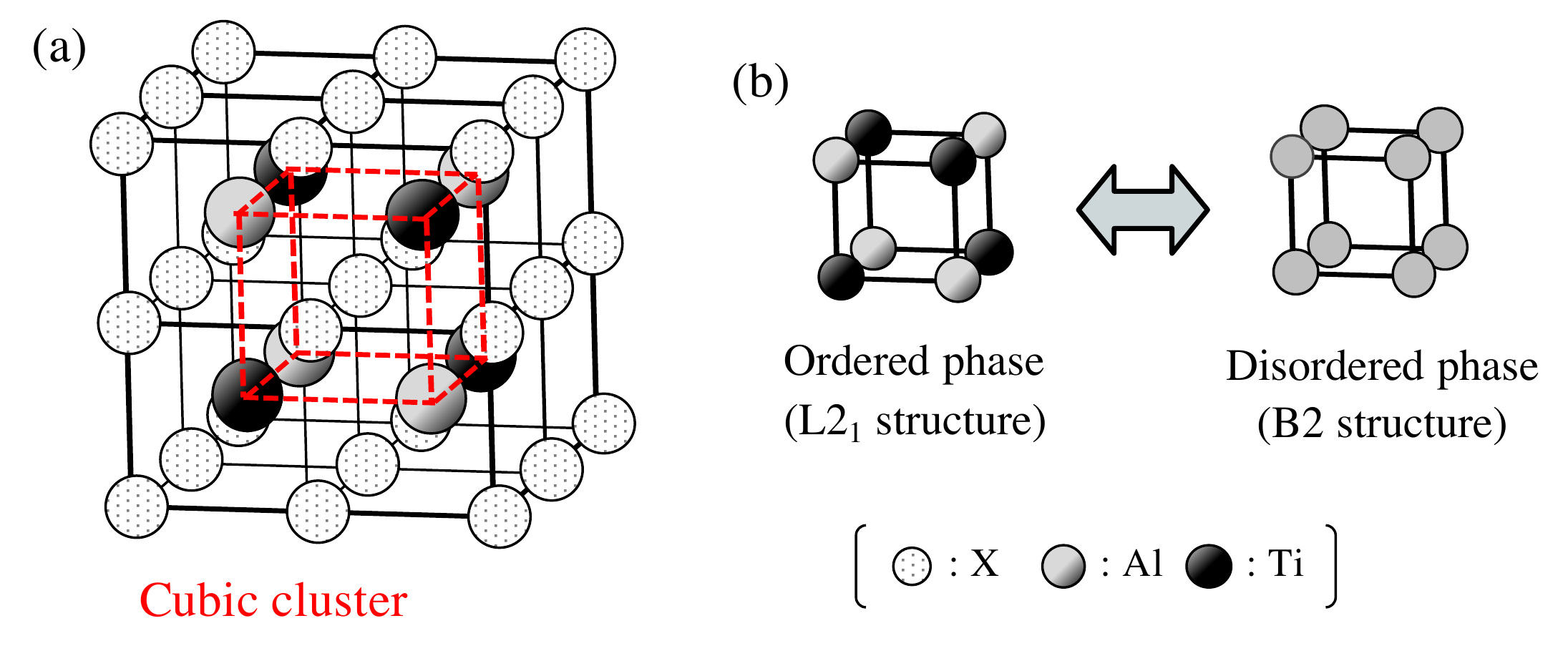}
\caption{\label{fig:B2_L21_phase_transformation} (a) L2$_1$ ordered structure, where X, Al, and Ti atoms are, respectively, shown in dotted, gray, and black balls. The lattice points which are predominately occupied by Al, Ti, and X atoms are, respectively, called the $\alpha$-, $\beta$-, and $\gamma$-sublattices. (b) The L2$_1$ structure transforms into the B2 structure when the Al and Ti atoms are randomly distributed on the $\alpha$- and $\beta$-sublattices, which can be viewed as an ordered--disordered phase transformation. The cubic cluster on the simple cubic lattice composed of the $\alpha$- and $\beta$-sublattices is shown in red, which is used as the basic cluster in the cubic approximation of the CVM. }
\end{figure}

On the other hand, since the transformation between the B2 and L2$_1$ phases in the X--Al--Ti alloys can be viewed as an order--disorder phase transformation (because the L2$_1$ phase is crystallographically equivalent to the B2 phase when the Al and Ti atoms on the $\alpha$- and $\beta$-sublattices are randomly distributed each other, as can be seen in Fig.\;\ref{fig:B2_L21_phase_transformation}\;(b)), the same reasoning employed for the Fe--Be alloy \cite{ino1978pairwise} is applicable. A systematic study has been conducted for X--Al--Ti alloys to investigate an effect of X (X=Fe, Co, Ni, and Cu) atoms on the stability of two-phase region, and a certain relationship between the phase separation and the number of 3d+4s valence electrons of the component elements has been proposed \cite{ishikawa2002phase}. This indicates that the phase separation into B2 and L2$_1$ ordered phases in the X--Al--Ti alloys is strongly related to chemical affinities. 

 An origin of the phase separation in the X--Al--Ti alloys is still controversial and has not been made clear. In this work, therefore, an origin of the phase separation in the Co-- and Ni--Al--Ti alloys as well as the reason of its absence in the Fe--Al--Ti alloy are explored using the cluster variation method (CVM) \cite{kikuchi1951theory}. CVM is one of the most reliable mean field approximations to formulate a free energy, and by combining it with an electronic structure total energy calculations (first-principles CVM \cite{mohri2013first}) it is expected that phase stability of the B2 and L2$_1$ ordered structures can be reliably evaluated in terms of both mechanical instability and chemical repulsions of unlike pairs. 
  
The paper is organized as follows. The methodology of an analysis of the phase stability of the B2 and L2$_1$ ordered structures using the first-principles CVM is described in Sec.\;II. In Section\;III, the calculated phase diagrams of the X--Al--Ti (X: Fe, Co, and Ni) alloys at the XAl--XTi pseudo-binary section are shown, and each contribution of mechanical instability and chemical repulsions is discussed. Finally, an origin of phase separation in the X--Al--Ti alloys is summarized in Sec.\;IV.

\section{\label{sec:level2}Theory}
The phase stability of the B2 and L2$_1$ ordered phases in the Ni--Al--Ti alloy have been analyzed using the CVM by Enomoto $et$ $al.$ \cite{enomoto1997analysis}, where the phenomenological interaction energies between the 1st- and 2nd-nearest-neighbor pairs on the original body-centered cubic (bcc) lattice were used with the tetrahedron approximation. The two-phase region of the B2 and L2$_1$ ordered phases has been successfully predicted in a Ti-rich side, and the phase separation behavior was explained in terms of the large lattice misfit between the B2 and L2$_1$ ordered phases without referring to any chemical affinities of atomic bonds in the system. In fact, to elucidate the chemical contribution, it is required to consider longer interaction energies than the 2nd-nearest-neighbor pairs on the bcc lattice (which corresponds to the 1st-nearest-neighbor pairs on the simple cubic lattice, see Fig.\;\ref{fig:B2_L21_phase_transformation}). A longer pair interaction energy cannot be taken into account in the tetrahedron approximation, and a higher order approximation needs to be employed to incorporate a longer pair interaction energy. The next higher order approximation for the bcc structure is the cube-octahedron approximation \cite{moran2012theory}, but the use of the cube-octahedron approximation is not so trivial because of its huge computational burden. 

The phase stability of the B2 and L2$_1$ ordered structures has also been explored in X$_2$A$_{2-x}$B$_x$ ($0 \leq x \leq 2$) alloys using the cubic approximation of the CVM by Kiyokane \cite{kiyokane2012cluster}, where only an interchange between A and B atoms on the $\alpha$- and $\beta$-sublattices of the L2$_1$ ordered structure was considered with the fixed X atoms on the $\gamma$-sublattice (as seen in Fig.\;\ref{fig:B2_L21_phase_transformation}, where A and B atoms would correspond to Al and Ti atoms, respectively). The author conducted a model calculation by setting an arbitrary constant value for the 1st-nearest-neighbor pair interaction energies on the simple cubic lattice (not on the bcc lattice) and determined a phase boundary between B2 and L2$_1$ phases at the XA--XB pseudo-binary section. Since any volume dependence on the pair interaction energies and any further distant pair interaction energies above the 1st-nearest neighbors were not included, a two-phase region, B2+L2$_1$, was not produced. However, this scheme has an advantageous feature over the tetrahedron CVM employed by Enomoto $et$ $al.$ \cite{enomoto1997analysis}, for it is possible to include longer interaction energies than the 2nd-nearest-neighbor pairs on the bcc lattice (or the 1st-nearest-neighbor pairs on the simple cubic lattice) with a relatively small computational burden compared with the cube-octahedron approximation.

In the present calculations, therefore, the cubic approximation of the CVM is employed taking into account multi-body interaction energies inside a cubic cluster (which include up to the 3rd-nearest-neighbor pair interaction energies on the simple cubic lattice) determined from an electronic band structure calculation. In the cubic approximation employed here, the only configurations of Al and Ti atoms on $\alpha$- and $\beta$-sublattices are considered as was done in Ref.\;\cite{kiyokane2012cluster}. The phase equilibria between the B2 and L2$_1$ ordered phases in the X$_2$Al$_{2-x}$Ti$_x$ ($0 \leq x \leq 2$) alloys are explored, and the phase diagrams for the pseudo-binary section, XAl--XTi, are determined. 

Note that a similar idea used here (i.e., limiting one's attention to atomic configurations on the $\alpha$- and $\beta$-sublattices) has been employed in the calculation of the tetragonal--cubic phase transformation in ZrO$_2$ \cite{mohri2013first,yamada2019application}, where only displacements of oxygen atoms are considered with the fixed zirconium positions. Since there is clear (reliable) experimental evidence that the Fe, Co, and Ni atoms in the X--Al--Ti alloys are located primarily on the $\gamma$-sublattice \cite{kainuma1997ordering,ishikawa2002ordering,ishikawa2002phase}, the present assumption is reasonable as long as the temperature is not too high. 

For an analysis of phase stability, a free energy of a system needs to be formulated. The configurational entropy, $S$, within the cubic approximation of the CVM is given by \cite{kiyokane2010order}
\begin{equation}
\begin{split}
S= k_B \; &\mbox{ln} \Bigg[ \frac{1}{2}\left( \sum_i L(x^{\alpha}_i) + \sum_i L(x^{\beta}_i) \right) - 3 \sum_{i,j} L(y^{\alpha \beta}_{ij})  \\
 + 3 & \sum_{i,j,k,l} L(z^{\alpha \beta \alpha \beta}_{ijkl}) - \sum_{i,j,k,l,m,n,o,p} L(w^{\alpha \beta \alpha \beta \alpha \beta \alpha \beta}_{ijklmnop})  \Bigg]   \; , \label{eq:entropy_cubic}
\end{split}
\end{equation}
where $k_B$ is the Boltzmann constant, $x^{\alpha/\beta}_i$, $y^{\alpha\beta}_{ij}$, $z^{\alpha \beta \alpha \beta}_{ijkl}$, and $w^{\alpha \beta \alpha \beta \alpha \beta \alpha \beta}_{ijklmnop}$ are, respectively, the cluster probabilities of the point, pair, square, and cubic, and $L(x) \equiv x\;\mbox{ln}x -x$. The subscripts and superscripts of the cluster probabilities indicate an atomic species and sublattice(s), respectively (e.g., $z^{\alpha \beta \alpha \beta}_{\mbox{\scriptsize{AlAlAlTi}}}$ represents the probability of finding the square configuration, Al-Al-Al-Ti, on the $\alpha$-$\beta$-$\alpha$-$\beta$ sublattices).
 
The total energy, $E$, in a system is described, in terms of a cubic interaction energy, as 
\begin{equation}
\begin{split}
E & = \sum_{i,j,k,l,m,n,o,p}\epsilon_{ijklmnop} w^{\alpha \beta \alpha \beta \alpha \beta \alpha \beta}_{ijklmnop}   \; , \label{eq:total_energy}
\end{split}
\end{equation}
where $\epsilon_{ijklmnop}$ is the cubic interaction energies and $w^{\alpha \beta \alpha \beta \alpha \beta \alpha \beta}_{ijklmnop} $ is the cubic cluster probabilities as defined above. Although there are $2^8=256$ possible cubic configurations of Al and Ti atoms, the number is significantly reduced to 22 due to a symmetry of the cubic structure. The cubic configurations are numbered and shown in Table\;\ref{table:cubic_clusters}. Note that the perfect B2 and L2$_1$ ordered phases (or XAl, XTi, and X$_2$AlTi) correspond to the 1st, 2nd, and 22nd cubic configurations in Table\;\ref{table:cubic_clusters}, respectively. 

\begin{table}
\begin{center}
\caption{\label{table:cubic_clusters} 22 kinds of cubic configurations, each of which is numbered and their interaction energies are shown together as $\epsilon_{n}$. \\}
{\renewcommand\arraystretch{1.5}
\begin{tabular}{ c  c  c  c } \hline \hline 
\includegraphics[scale=0.42]{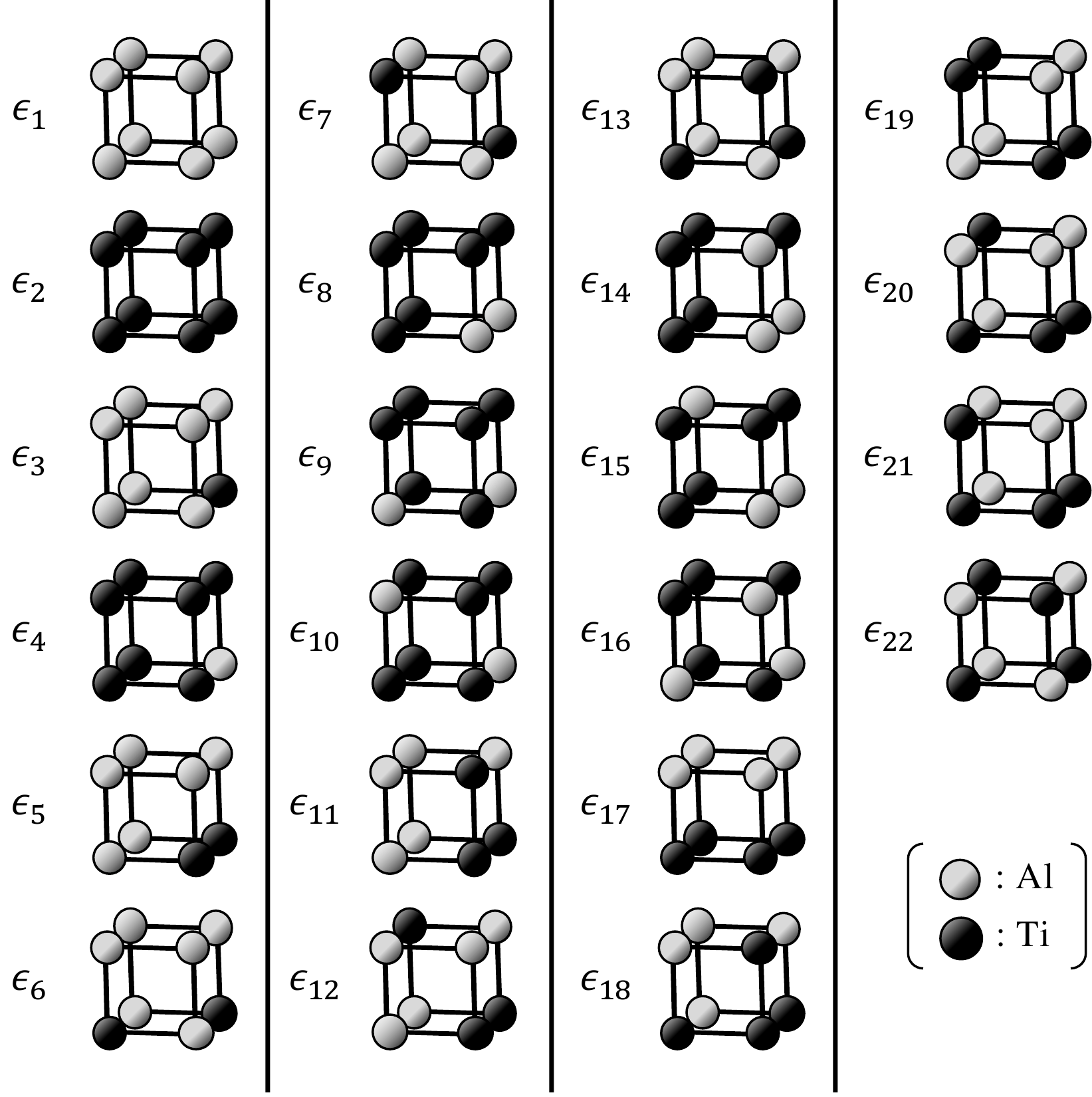}
 \\ \hline \hline
\end{tabular}
}
\end{center}
\end{table}

The cubic interaction energies, $\epsilon_{ijklmnop}$ (or $\epsilon_{n}$ where $n$ represents a kind of the cubic configurations shown in Table\;\ref{table:cubic_clusters}), are obtained from the electronic band structure calculation using the projector augmented-wave (PAW) method \cite{kresse1996efficiency} as implemented in the Vienna $Ab$ $Initio$ Simulation Package (VASP). A non-spin-polarization calculation is conducted with the generalized gradient approximation (GGA) of Perdew-Burke-Ernzerhof (PBE) \cite{perdew1996generalized} for the exchange-correlation functional. Supercells in the present DFT calculations contain 16 atoms (the number of X, Al, and Ti atoms are, respectively, 8, $8-N$, and $N$, where $N$ takes an integer value from 0 to 8) in the bcc structure, where X atoms are fixed on the $\gamma$-sublattice and Al and Ti atoms are placed on the $\alpha$- and $\beta$-sublattices as shown in Table\;\ref{table:cubic_clusters}. The plane wave cut-off energy is set to 400\;eV, and integration over the Brillouin zone is done with $4$$\times$$4$$\times$$4$ $k$-points. The cubic interaction energies calculated at various volumes are fitted to the following Murnaghan equation of state \cite{tyuterev2006murnaghan}:
\begin{equation}
\begin{split}
\epsilon_{ijklmnop} & (V)  = \epsilon_{ijklmnop} (V_0)  \\
+ & \frac{B_0 V}{B'_0 (B'_0-1)} \left[ B'_0 \left( 1-\frac{V_0}{V} \right) + \left( \frac{V_0}{V} \right)^{B'_0} -1 \right] \; ,  \label{Murnaghan_equation_of_state}
\end{split}
\end{equation}
where $V_0$ is an equilibrium volume at the ground state, $B_0$ is the bulk modulus at the equilibrium volume, and $B'_0$ is the first derivative of the bulk modulus, $B$, with respect to pressure, $P$, evaluated at the $V_0$; i.e., $B'_0=(\partial B / \partial P)_{V_0}$. 

An equilibrium state of a system at a finite temperature, $T$, is determined using the grand potential, $\Omega$. The grand potential is derived by performing the Legendre transformation on the Helmholtz free energy, $F$, in terms of point cluster probabilities, $x_i$, as $\Omega=F-\sum_i \mu_i x_i$, where $\mu_i$ is the chemical potential for the atomic species, $i$. The equilibrium state is determined by imposing the following two conditions:
\begin{equation}
\left( \frac{\partial \Omega}{\partial w^{\alpha \beta \alpha \beta \alpha \beta \alpha \beta}_{ijklmnop}} \right)_{V,T} = 0   \;  \label{eq:free_energy_minimization_cluster_variable}
\end{equation}
and
\begin{equation}
\left( \frac{\partial \Omega}{\partial V} \right)_{T, x_{i}^{\alpha}, x_{i}^{\beta}, y_{ij}^{\alpha \beta}, z_{ijkl}^{\alpha \beta \alpha \beta},  w^{\alpha \beta \alpha \beta \alpha \beta \alpha \beta}_{ijklmnop}} = 0   \; . \label{eq:free_energy_minimization_distance}
\end{equation}
The minimization of the grand potential in terms of cluster probability (Eq.\;\eqref{eq:free_energy_minimization_cluster_variable}) is conducted by the natural iteration method (NIM) \cite{kikuchi1974superposition}. The phase boundary between L2$_1$ (or ordered) and B2 (or disordered) phases at each temperature is determined from the condition that their equilibrium grand potentials, $\Omega_{eq}(T)$, become identical; i.e., $\Omega_{eq}^{L2_1}(T)=\Omega_{eq}^{B2}(T)$.

\section{\label{sec:level3}Results and Discussion}
\begin{figure}
\includegraphics[scale=0.47]{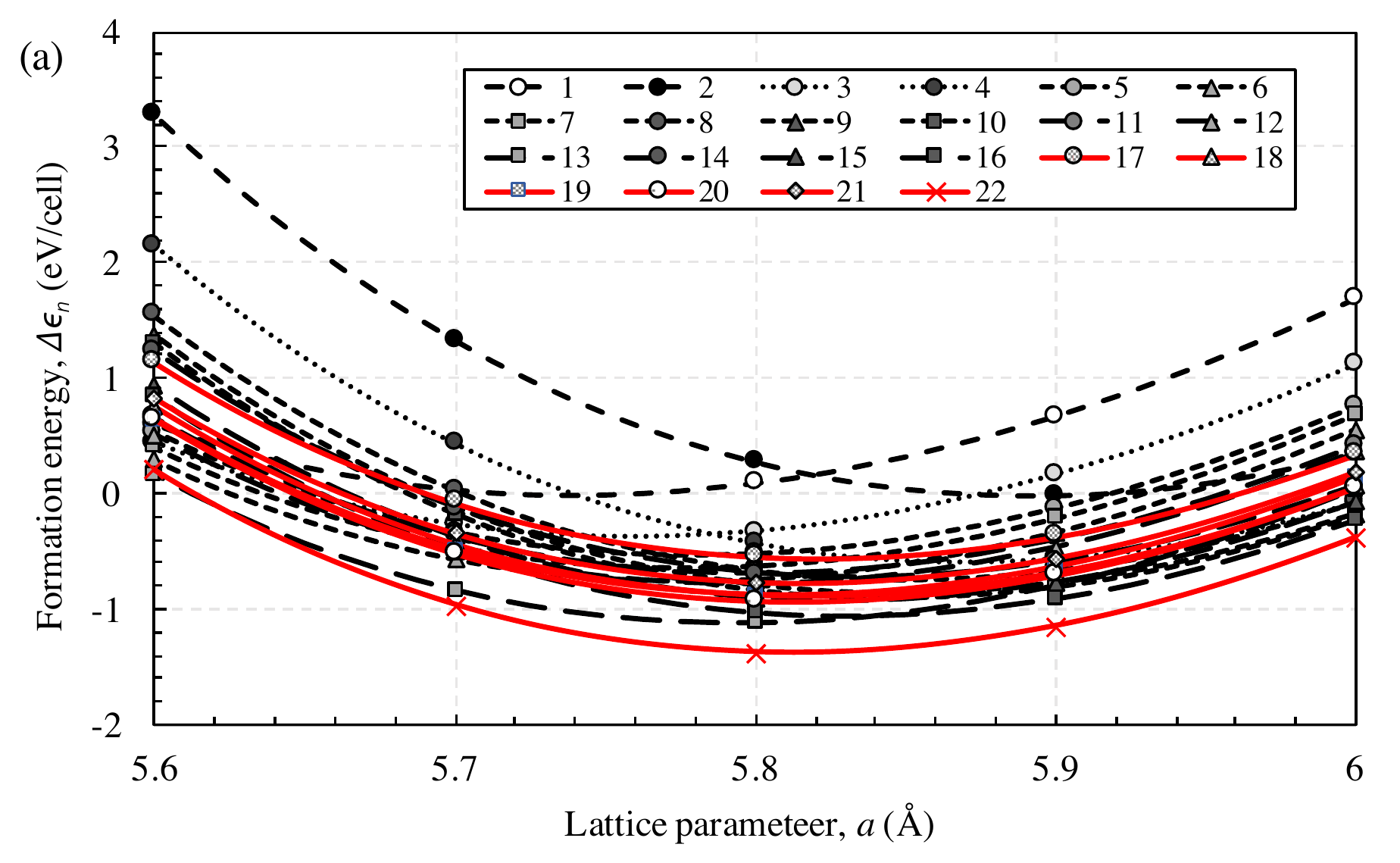}
\includegraphics[scale=0.47]{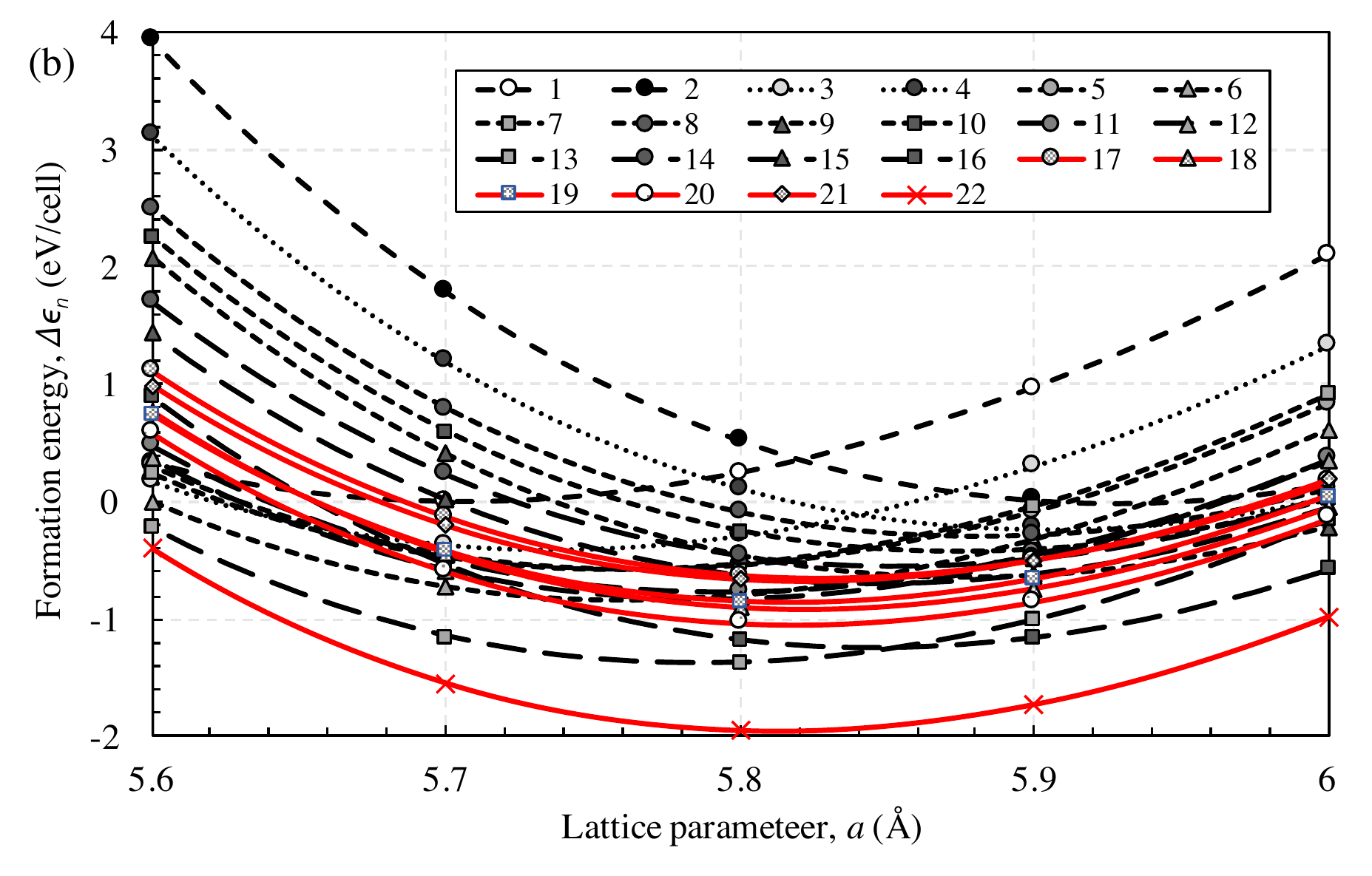}
\includegraphics[scale=0.47]{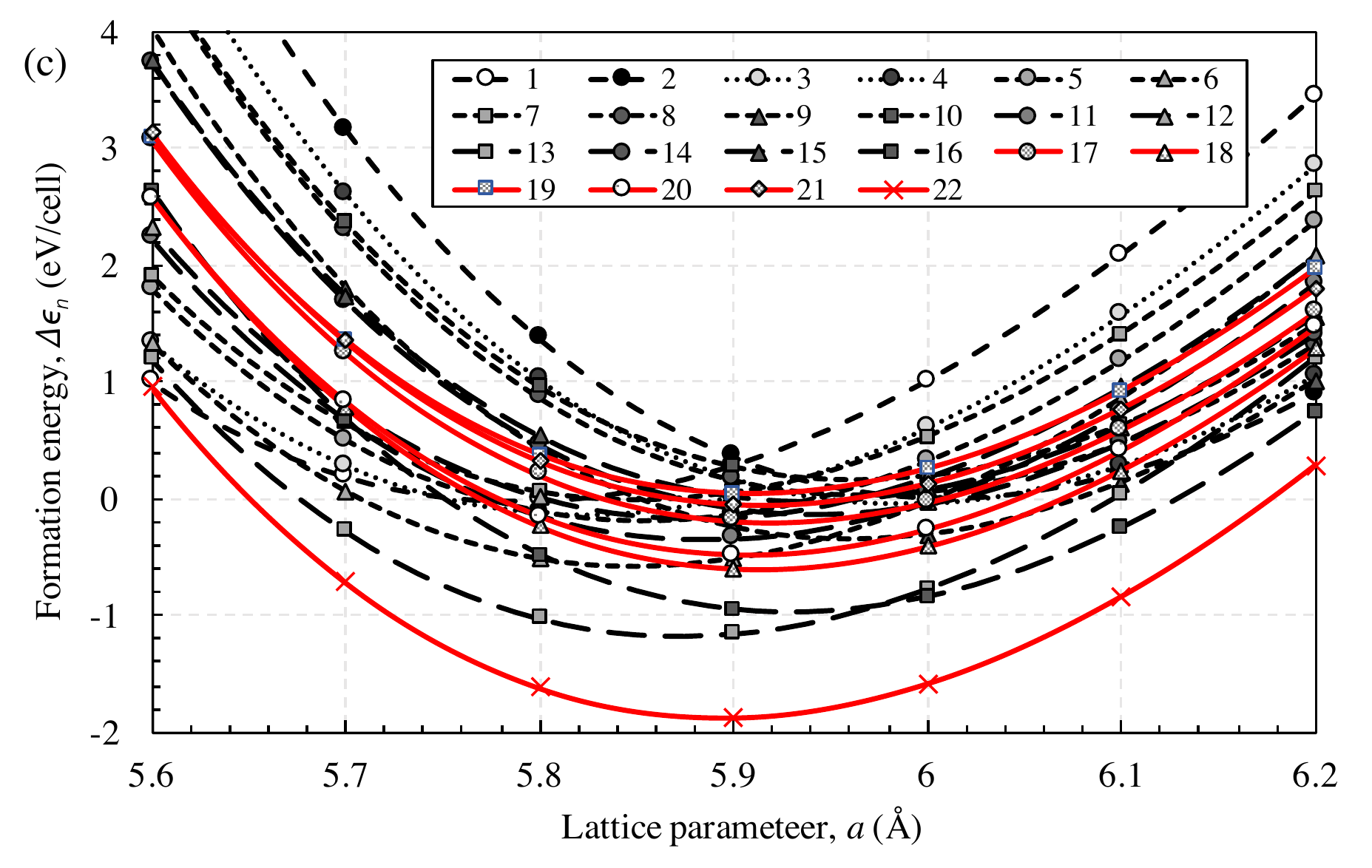}
\caption{\label{fig:formation_energy} Calculated formation energies of the atomic configurations shown in Table\;\ref{table:cubic_clusters} for the (a) Fe--Al--Ti, (b) Co--Al--Ti, and (c) Ni--Al--Ti alloys. Here, the segregation limit connecting the minimum energies of the perfect B2 ordered phases, XAl and XTi, is taken as the reference energy.  }
\end{figure}

The calculated formation energies, $\Delta \epsilon_n$ (or $\Delta \epsilon_{ijklmnop}$), of each atomic configuration in the X--Al--Ti (X: Fe, Co, and Ni) alloys in terms of a lattice parameter, $a$ ($=V^{\frac{1}{3}}$), are shown in Fig.\;\ref{fig:formation_energy}, where the segregation limit connecting the minimum energies of the perfect B2 ordered phases, XAl and XTi, is taken as the reference energy. From Fig.\;\ref{fig:formation_energy}, it can be seen that the L2$_1$ ordered structure has the minimum formation energy in the all Fe--, Co--, and Ni--Al--Ti alloys. This indicates that the L2$_1$ ordered structure is the most stable phase at the ground state. The equilibrium lattice constant, $a_0$, and bulk modulus, $B_0$, of the perfect B2 and L2$_1$ ordered phases determined from the Murnaghan equation of state, Eq.\;\eqref{Murnaghan_equation_of_state}, are summarized in Table\;\ref{table:lattice_constant_bulk_modulus}. It is found that the calculated lattice constants are close to the experimental data, but there are nontrivial differences in the bulk moduli. 

\begin{table}
\begin{center}
\caption{\label{table:lattice_constant_bulk_modulus}Equilibrium lattice constant, $a_0$, and bulk modulus, $B_0$, of the perfect B2 and L2$_1$ ordered phases at the ground state evaluated using the Murnaghan equation of state, Eq.\;\eqref{Murnaghan_equation_of_state}. Their available experimental data are shown together (note that the experimental data were measured at ambient temperatures). \\}
\footnotesize
{\renewcommand\arraystretch{1.5}
\begin{tabular}{ c  c  c  c  } \hline \hline
 $a_0$ ($\mathrm{\AA}$)  & \quad \quad Cal. \quad \quad & \quad \quad Exp. \quad \quad  & \quad Diff. (\%) \quad  \\ \hline
FeAl (B2)  & 2.869 & 2.862 \cite{villars1986pearson}  &  +0.24 \\
FeTi (B2)  & 2.947 & 2.972 \cite{van198657fe}  &  -0.85  \\
Fe$_2$AlTi (L2$_1$)  & 5.812 & 5.882 \cite{okpalugo1985onset}   &  -1.20 \\
CoAl (B2)  & 2.851 & 2.861 \cite{ogut1994ab}  &  -0.25 \\
CoTi (B2)  & 2.965 & 2.995 \cite{villars1986pearson}  &  -1.01  \\
Co$_2$AlTi (L2$_1$)  & 5.812 &  5.85 \cite{carbonari1996magnetic}   & -0.65 \\
NiAl (B2)  & 2.894 & 2.886 \cite{otto1997equation}  &  +0.28 \\
NiTi (B2)  & 3.006 & 3.012 \cite{matsumoto1989lattice}  &  -0.20  \\
Ni$_2$AlTi (L2$_1$)  & 5.896 & 5.90 \cite{oh1997microstructure}  &  -0.40  \\ \hline \hline
 $B_0$ ($\mathrm{GPa}$)  & \quad \quad Cal. \quad \quad & \quad \quad Exp. \quad \quad  & \quad Diff. (\%) \quad \\ \hline
FeAl (B2)  & 176.8 &  152 \cite{zhang1995analysis} &  +16.3 \\
FeTi (B2)  & 194.0 &  160.8 \cite{liebertz1980growth}, 189 \cite{buchenau1983lattice}  &  +17.1, +2.64  \\
Fe$_2$AlTi (L2$_1$)  & 187.0 &  --  &  -- \\
CoAl (B2)  & 175.7 & 162 \cite{ogut1994ab}  &  +7.80 \\
CoTi (B2)  & 181.0 & 154 \cite{yasuda1991elastic}  &  +16.6  \\
Co$_2$AlTi (L2$_1$)  & 183.3 &  --  &  -- \\
NiAl (B2)  & 155.1 & 156 \cite{otto1997equation} &  -0.58 \\
NiTi (B2)  & 163.7 &  140.3 \cite{mercier1980single}  &  +14.3  \\
Ni$_2$AlTi (L2$_1$)  & 162.7 & --  &  -- \\  \hline \hline
\end{tabular}
}
\end{center}
\end{table}

\begin{figure}
\includegraphics[scale=0.56]{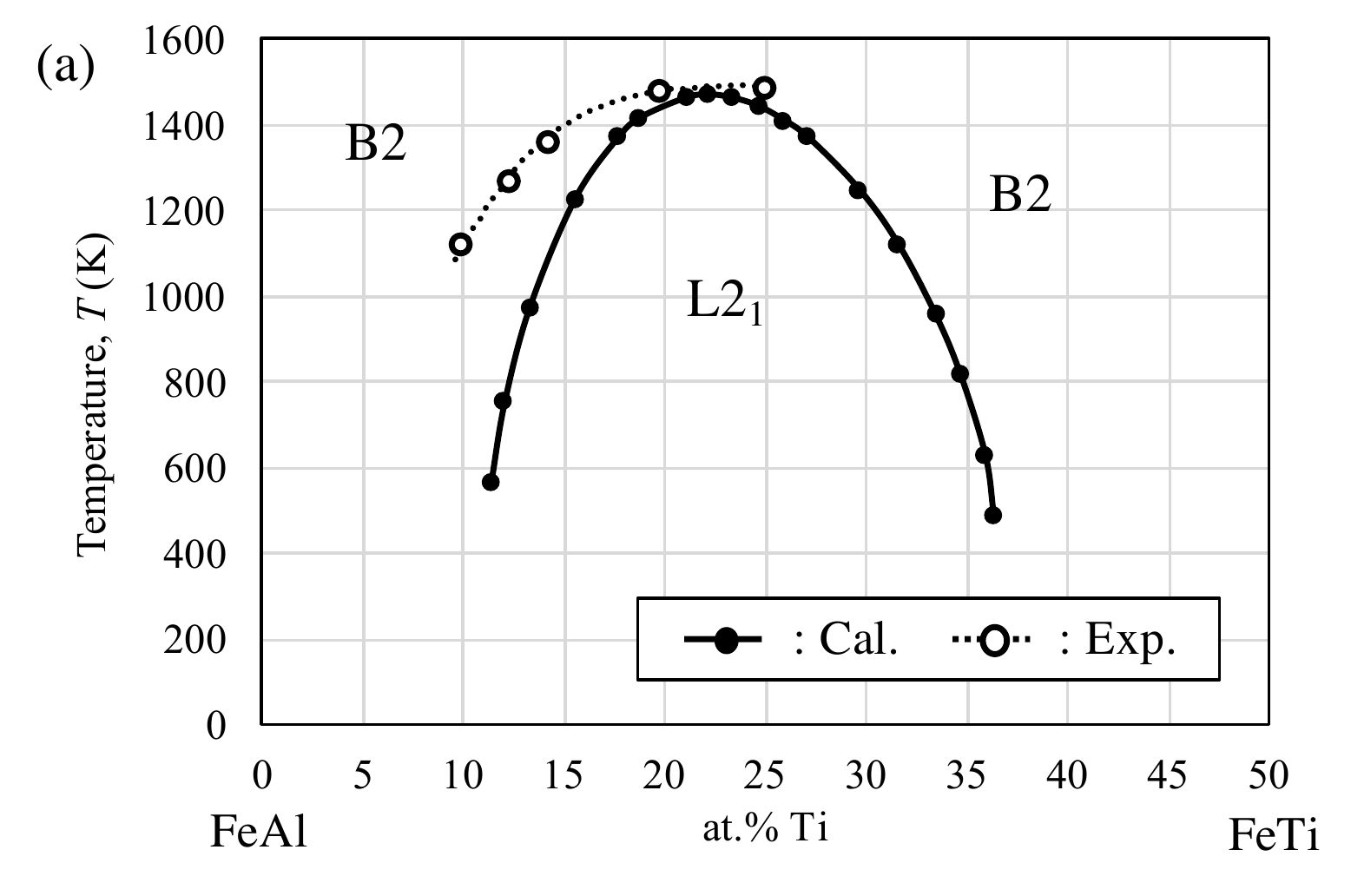}
\includegraphics[scale=0.56]{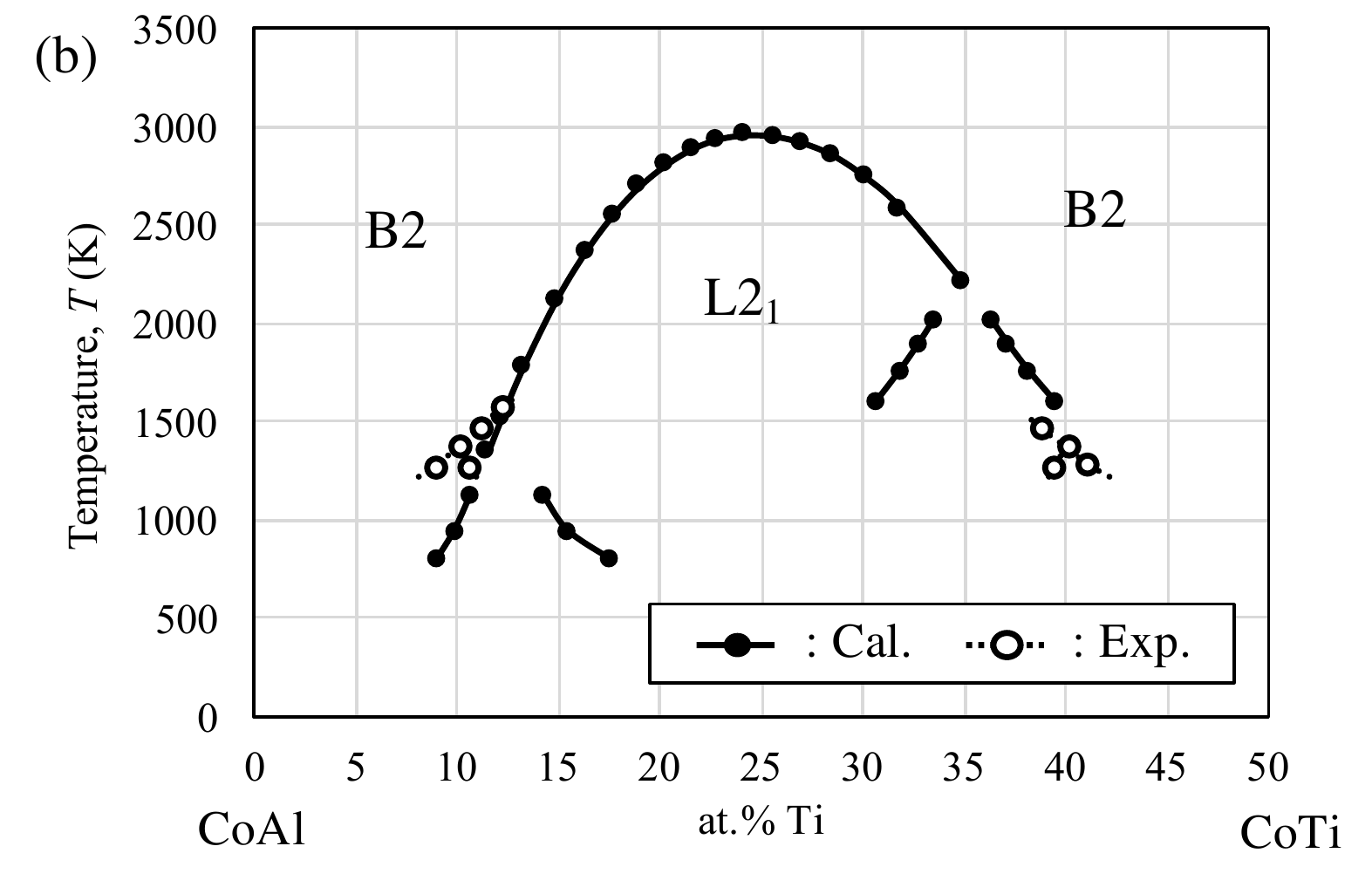}
\includegraphics[scale=0.56]{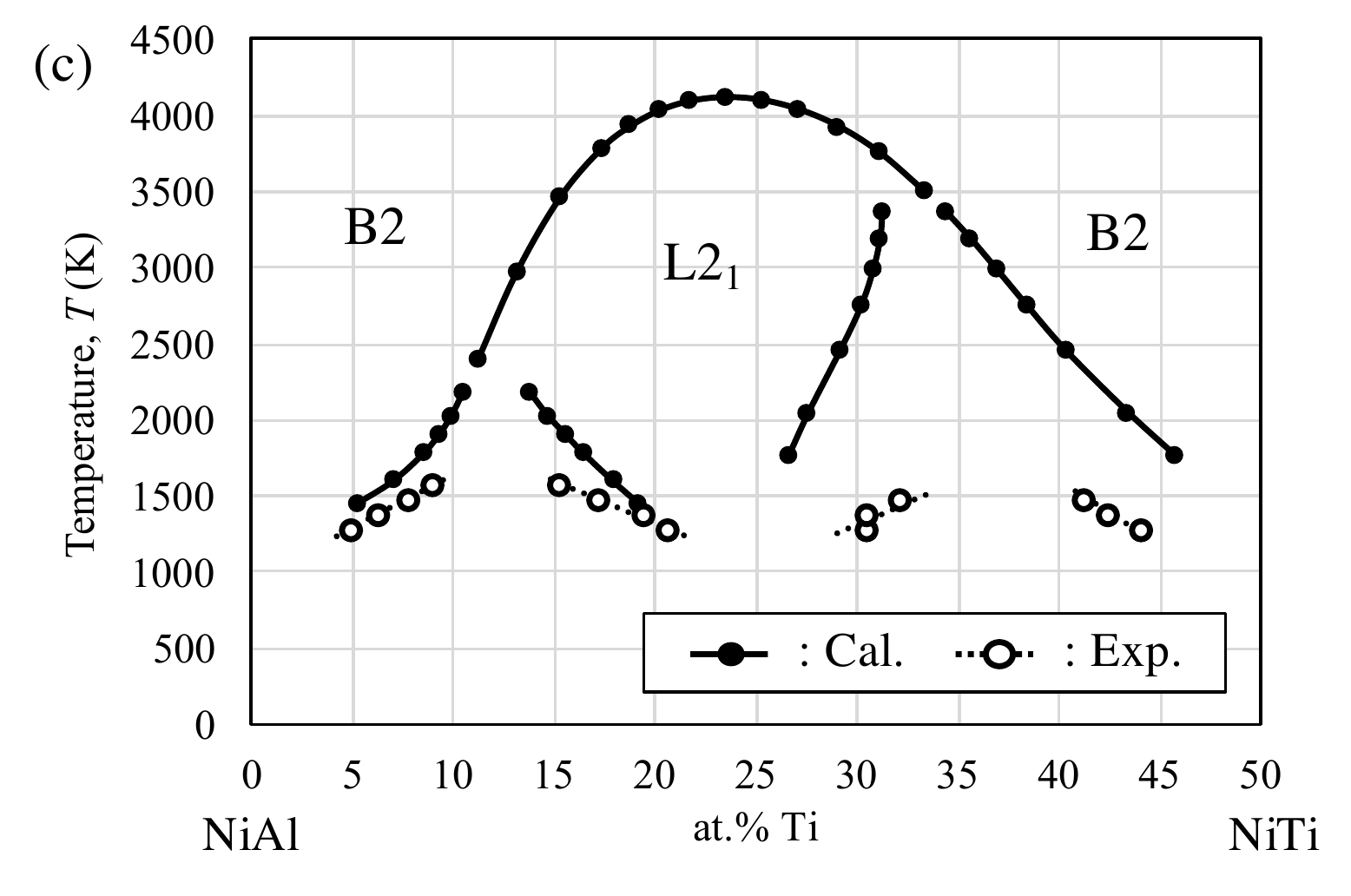}
\caption{\label{fig:phase_diagram} Phase boundaries between the B2 and L2$_1$ ordered phases in the (a) Fe--Al--Ti, (b) Co--Al--Ti, and (c) Ni--Al--Ti alloys at the pseudo-binary section, XAl--XTi (X: Fe, Co, and Ni). The calculated results and experimental data \cite{kainuma1997ordering,ishikawa2002ordering,ishikawa2002phase} are, respectively, shown as filled and open circles. }
\end{figure}

The calculated phase diagrams in the X--Al--Ti alloys at the pseudo-binary section, XAl--XTi, are shown in Fig.\;\ref{fig:phase_diagram}, where their experimental results \cite{kainuma1997ordering,ishikawa2002ordering,ishikawa2002phase} are shown together. It can be seen that there are only single phase regions of B2 and L2$_1$ ordered phases in the Fe--Al--Ti alloy, whereas there are two-phase regions at low temperatures in both Co--Al--Ti and Ni--Al--Ti alloys. These are consistent with the experimental results \cite{kainuma1997ordering,ishikawa2002ordering,ishikawa2002phase}.

\begin{figure}
\includegraphics[scale=0.10]{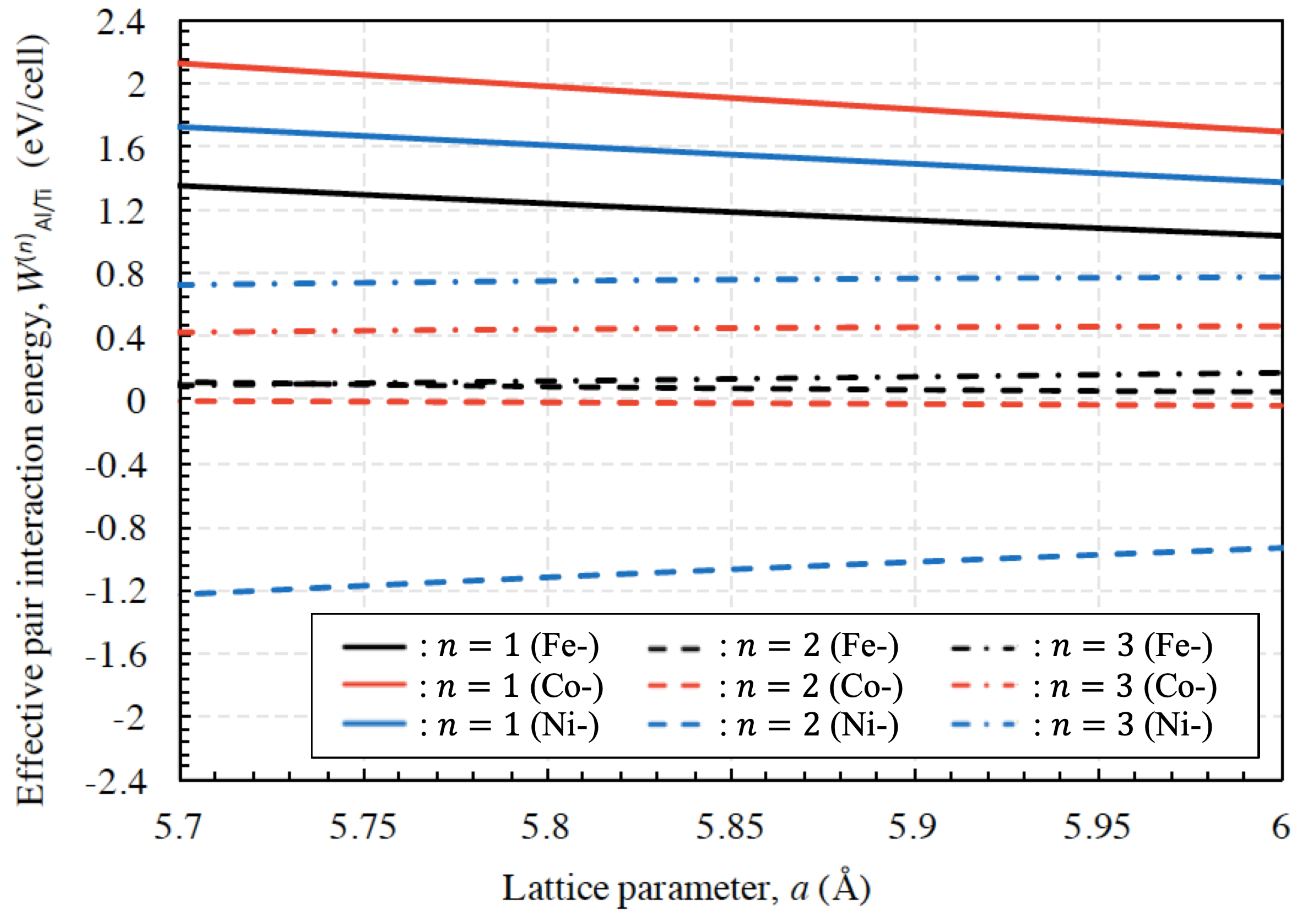}
\caption{\label{fig:effective_pair_interaction_energy} Effective pair interaction energies of the $n$th-nearest-neighbor pairs, $W_{\mbox{\scriptsize{Al/Ti}}}^{(n)}$, calculated using the CEM. The black, red, and blue lines are, respectively, for the Fe--, Co--, and Ni--Al--Ti alloys. The 1st, 2nd, and 3rd effective interactions are shown in solid, broken, and broken-dotted lines, respectively. }
\end{figure}

To clarify the reason of the absence of the phase separation in the Fe--Al--Ti alloy, the lattice misfits between the B2 and L2$_1$ ordered phases, $\delta_{B2/L2_1}$, at the ground state and the 1st- to 3rd-nearest-neighbor effective pair interaction energies on the simple cubic lattice, $W_{\mbox{\scriptsize{Al/Ti}}}^{(\mbox{\scriptsize{1,2, or 3}})}$, are calculated in the X--Al--Ti alloys and shown in Table\;\ref{table:lattice_misfit} and Fig.\;\ref{fig:effective_pair_interaction_energy}, respectively. The lattice misfit is defined as
\begin{equation}
\delta_{B2/L2_1}=\frac{| a^{L2_1}_0- a^{B2}_0 |}{a^{B2}_0}   \; , \label{eq:lattice_misfit}
\end{equation}
where $a^{B2}_0$ and $a^{L2_1}_0$ are the lattice constants of the B2 and L2$_1$ ordered phases at the ground state, respectively. The effective pair interaction energies are defined as $W_{\mbox{\scriptsize{Al/Ti}}}^{(n)}\equiv e_{\mbox{\scriptsize{AlAl}}}^{(n)}+e_{\mbox{\scriptsize{TiTi}}}^{(n)}-2e_{\mbox{\scriptsize{AlTi}}}^{(n)}$, where $e_{\mbox{\scriptsize{AlAl}}}^{(n)}$, $e_{\mbox{\scriptsize{AlTi}}}^{(n)}$, $e_{\mbox{\scriptsize{TiTi}}}^{(n)}$ are, respectively, the $n$th-nearest-neighbor pair interaction energies of Al--Al, Al--Ti, and Ti--Ti pairs on the simple cubic lattice. The effective pair interaction energies can be used to evaluate a tendency of ordering or phase separation in a system: when $W_{\mbox{\scriptsize{Al/Ti}}}^{(n)}>0$ ($W_{\mbox{\scriptsize{Al/Ti}}}^{(n)}<0$) the Al--Al and Ti--Ti pairs (Al--Ti and Ti--Al pairs) are preferred in the $n$th-nearest-neighbors. These are extracted using the cluster expansion method (CEM) \cite{connolly1983density} (the detail of the CEM can be found in Appendix\;\ref{level5_1}). From Table\;\ref{table:lattice_misfit} and Fig.\;\ref{fig:effective_pair_interaction_energy}, it is found that the lattice misfits for the Fe--Al--Ti alloy are much smaller than those for the Co-- and Ni--Al--Ti alloys. Furthermore, the 1st-nearest-neighbor interaction energy in the Fe--Al--Ti alloy is dominant compared to the 2nd- and 3rd-nearest-neighbors. These indicate that it is difficult to induce neither mechanically nor chemically driven phase separation in the Fe--Al--Ti alloy, so that any phase separation behavior, or two-phase region, could not be produced.

\begin{table}
\begin{center}
\caption{\label{table:lattice_misfit}Lattice misfits between the B2 and L2$_1$ ordered structures at the ground state, which are calculated from Table\;\ref{table:lattice_constant_bulk_modulus} and Eq.\;\eqref{eq:lattice_misfit}.  \\}
\footnotesize
{\renewcommand\arraystretch{1.5}
\begin{tabular}{ c  c  c  c  } \hline \hline
 $\delta_{B2/L2_1}$  & \quad  Fe--Al--Ti  \quad & \quad  Co--Al--Ti  \quad  & \quad  Ni--Al--Ti  \quad  \\ \hline
$\delta_{\mbox{\scriptsize{XAl/X2AlTi}}}$  & 1.288 $\%$ & 1.922 $\%$  &  1.875 $\%$ \\
$\delta_{\mbox{\scriptsize{XTi/X2AlTi}}}$  & 1.398 $\%$ & 1.991 $\%$   &  1.942 $\%$  \\ \hline \hline
\end{tabular}
}
\end{center}
\end{table}

\begin{figure}
\includegraphics[scale=0.61]{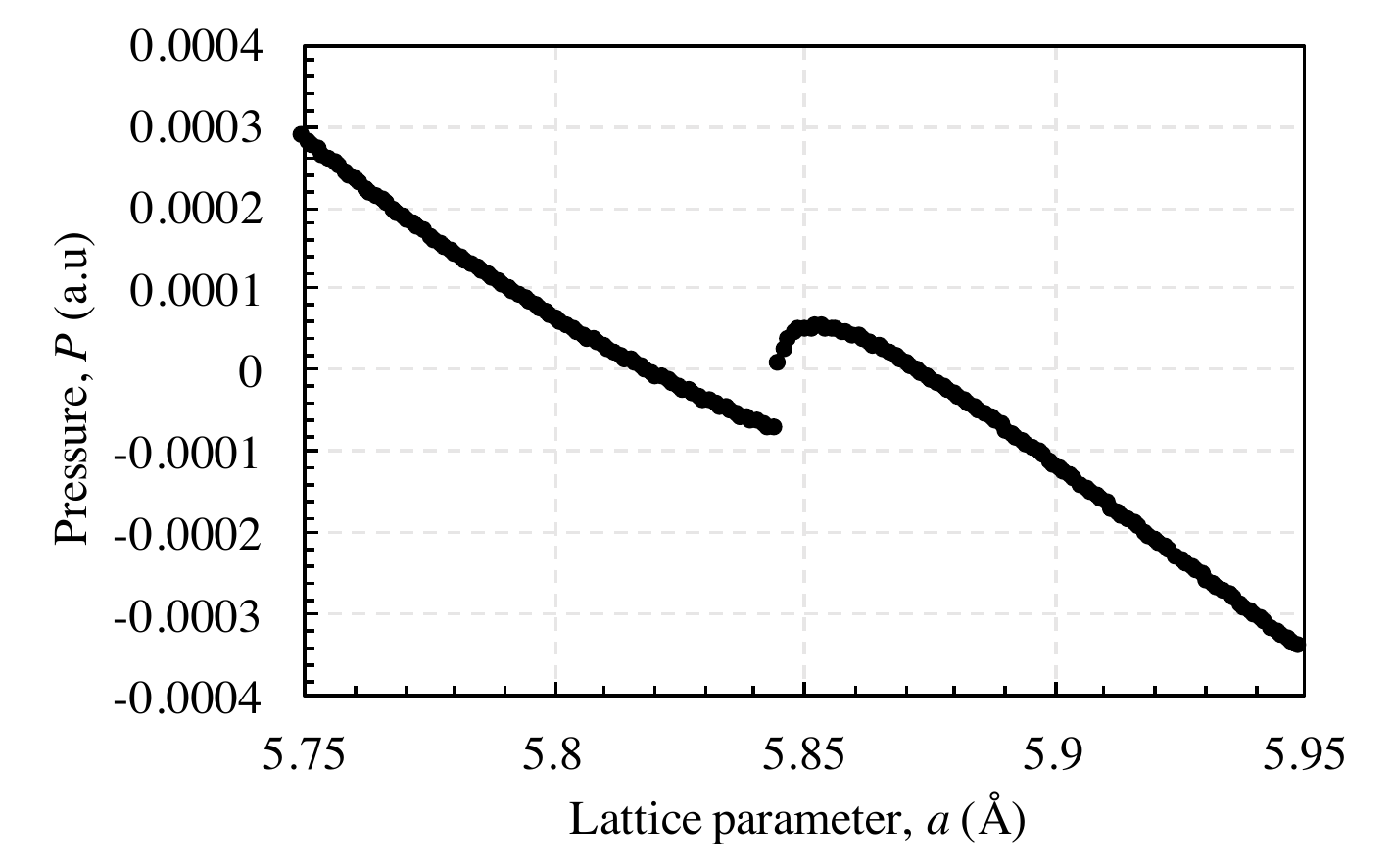}
\caption{\label{fig:p_v_curve_Ni_Al_Ti} $P$--$V$ curve at the Al--rich side of the Ni--Al--Ti alloy at $T=1500$\;K. }
\end{figure}

In the Ni--Al--Ti alloy, on the contrary, the lattice misfit is quite large and the 2nd-nearest-neighbor effective pair interaction shows a large negative value from Table\;\ref{table:lattice_misfit} and Fig.\;\ref{fig:effective_pair_interaction_energy}. These make the system mechanically unstable and chemically frustrating, which result in the phase separation. To confirm the mechanical instability of the Ni--Al--Ti alloy, the $P$--$V$ curves are calculated. One of the representative $P$--$V$ curves in the Ni--Al--Ti alloy is shown in Fig\;\ref{fig:p_v_curve_Ni_Al_Ti}, which is the one at $T=1500$\;K in the Al--rich side. It can be seen that the curve intersects with the horizontal axis, or $P=0$, at three different lattice parameters. Among these, the left and right intersection points satisfy the stability criteria, $\partial P/\partial V < 0$, and are considered to be stable phases, of which lattice constants are given at intersection points. The middle intersection point, on the other hand, is unstable from the stability criteria, and it decomposes into the two stable phases. It was confirmed that the lattice parameters at the two intersections are the same with those of the B2 and L2$_1$ ordered phases independently calculated in Fig.\;\ref{fig:phase_diagram}.

For the Co--Al--Ti alloy, the lattice misfit is as large as that for the Ni--Al--Ti alloy, but the 2nd- and 3rd-nearest-neighbor effective interaction energies are not significant compared to the 1st-nearest-neighbors. Therefore, it is considered that the phase separation behavior in the Co--Al--Ti alloy is purely due to the mechanical instability. This explains the smaller two-phase regions of B2 and L2$_1$ ordered phases in the Co--Al--Ti alloy than those in the Ni--Al--Ti alloy (see Fig.\;\ref{fig:phase_diagram}\;(b) and (c)).   

The negatively large effective pair interaction energy of the 2nd-nearest-neighbor pairs in the Ni--Al--Ti alloy is associated with the positive formation energies of some atomic configurations (see Fig.\;\ref{fig:formation_energy}\;(c)).
In the Fe-- and Co--Al--Ti alloys, on the contrary, there are few configurations that have a positive formation energy (as can be seen in Fig.\;\ref{fig:formation_energy}\;(a) and (b)), resulting in a positive and very small effective pair interaction energy in the Fe-- and Co--Al--Ti alloys, respectively (see Fig.\;\ref{fig:effective_pair_interaction_energy}).

Note that there are some deviations in the calculated phase diagrams compared with the experiments, such as a smaller B2 phase region in the Fe--Al--Ti alloy and an overestimation of the two-phase regions in the Ti-rich side in the Co-- and Ni--Al--Ti alloys. There are several possible reasons for the deviations, such as an ignorance of local atomic displacements, an assumption that X atoms are located only at the $\gamma$-sublattice, and an accuracy of the electronic band structure calculations. For a further reliable description of the phase diagrams, it is required to cope with these issues. As for the overestimation of two-phase regions mentioned above, we believe that an accuracy of the interaction energies determined from the band structure calculations is a main problem. As can be seen in Table\;\ref{table:lattice_constant_bulk_modulus}, the deviations of bulk modulus are much larger in the Ti-rich side than those in the Al-rich side.

\section{\label{sec:level4}Conclusions}
The phase diagrams of the X--Al--Ti (X: Fe, Co, and Ni) alloys at the XAl--XTi pseudo-binary section are calculated by the CVM using the interaction energies determined from the electronic band structure calculations. The cubic approximation is employed assuming only an interchange between the Al and Ti atoms on the $\alpha$- and $\beta$-sublattices with the X atoms fixed on the $\gamma$-sublattice. The special attention is paid to the stability of B2 and L2$_1$ ordered phases as well as an origin of the phase separation behaviors in these alloy systems. 

The calculated phase diagrams show that there are only single phase regions in the Fe--Al--Ti alloy, whereas there are two-phase regions of B2 and L2$_1$ ordered structures in both Al- and Ti-rich sides in the Co-- and Ni--Al--Ti alloys. From the lattice misfits between B2 and L2$_1$ phases and the effective pair interaction energies, it is found that since neither mechanical instability nor chemical repulsions of unlike pairs are expected in the Fe--Al--Ti alloy, no phase separation behavior is observed. On the other hand, mechanical instability and both mechanical instability and chemical repulsions are expected in the Co-- and Ni--Al--Ti alloy, respectively. Thus, the phase separation behaviors in the Co-- and Ni--Al--Ti alloy systems are, respectively, ascribed to the mechanical instability and the combination of mechanical instability and chemical repulsions.

It is believed that the present formalism can be applied to other metallic alloy systems in which a phase separation has been observed. Clarifying an origin of phase separation in various alloy systems will permit one to control a microstructure of industrial materials in an effective way.  \\

\section*{Acknowledgement}
We acknowledge Dr. K. Sato at Division of Materials and Manufacturing Science, Graduate School of Engineering, Osaka University, Suita, Osaka for providing computational resources that have contributed to the results reported within this paper.

\begin{appendices}

\section*{Appendix}
\section{\label{level5_1}Cluster expansion method}
Cluster expansion method (CEM) \cite{connolly1983density} is employed to extract the effective pair interaction energies of the 1st-, 2nd-, and 3rd-nearest-neighbors between the Al and Ti atoms on the simple cubic lattice. With the use of the CEM, the formation energies, $\Delta E^{(n)}$(which corresponds to $\Delta \epsilon_{n}$ in this work, where $n$ denotes a kind of the cubic configurations shown in Table\;\ref{table:cubic_clusters}), are written as
\begin{equation}
 \Delta  E^{(n)} = \sum_{m} v_m \xi_m^{(n)}  \; ,  \label{eq:CEM_original}
\end{equation}
where $v_m$ is the effective cluster interaction energy for a cluster $m$ and $\xi_m^{(n)}$ is a correlation function. The cubic cluster is composed of 20 sub-clusters \cite{kiyokane2012cluster,kiyokane2010order} as shown in Table\;\ref{table:subclusters}. The correlation functions, $\xi_m^{(n)}$, are uniquely determined for the each cubic configuration, $n$, using a spin operator $\sigma(p)$, which takes either $+1$ or $-1$ depending upon the Al or Ti atom located at a lattice site $p$. Using the $\xi_m^{(n)}$ with the $ \Delta  E^{(n)}$ calculated from the band calculations (shown in Fig.\;\ref{fig:formation_energy}), the effective cluster interaction energies, $v_m$, are determined as
\begin{equation}
 v_m = \sum_{n} \left( \xi_m^{(n)} \right)^{-1} \Delta E^{(n)}   \; .  \label{eq:CEM_inversion}
\end{equation}
The $v_2$, $v_3$, and $v_4$ correspond to the 1st-, 2nd-, and 3rd-nearest neighbor effective pair interaction energies, and they are related with the $W_{\mbox{\scriptsize{Al/Ti}}}^{(n)}$ as $W_{\mbox{\scriptsize{Al/Ti}}}^{(1)}=2 v_2$, $W_{\mbox{\scriptsize{Al/Ti}}}^{(2)}=2 v_3$, and $W_{\mbox{\scriptsize{Al/Ti}}}^{(3)}=2 v_4$.  

\begin{table}
\begin{center}
\caption{\label{table:subclusters} 20 sub-clusters in the cubic cluster \cite{kiyokane2012cluster,kiyokane2010order}. Here, the cubic cluster is shown as well. The each sub-cluster is numbered, and the corresponding correlation function is shown together. The $\xi_1$, $\xi_{2-4}$, $\xi_{5-7}$, $\xi_{8-13}$, $\xi_{14-16}$, $\xi_{17-19}$, $\xi_{20}$, and $\xi_{21}$ are, respectively, the point, pair, triangle, four-body, five-body, six-body, seven-body, eight-body (or cubic) correlation functions.  \\}
{\renewcommand\arraystretch{1.5}
\begin{tabular}{ c  c  c  c } \hline \hline 
\includegraphics[scale=0.6]{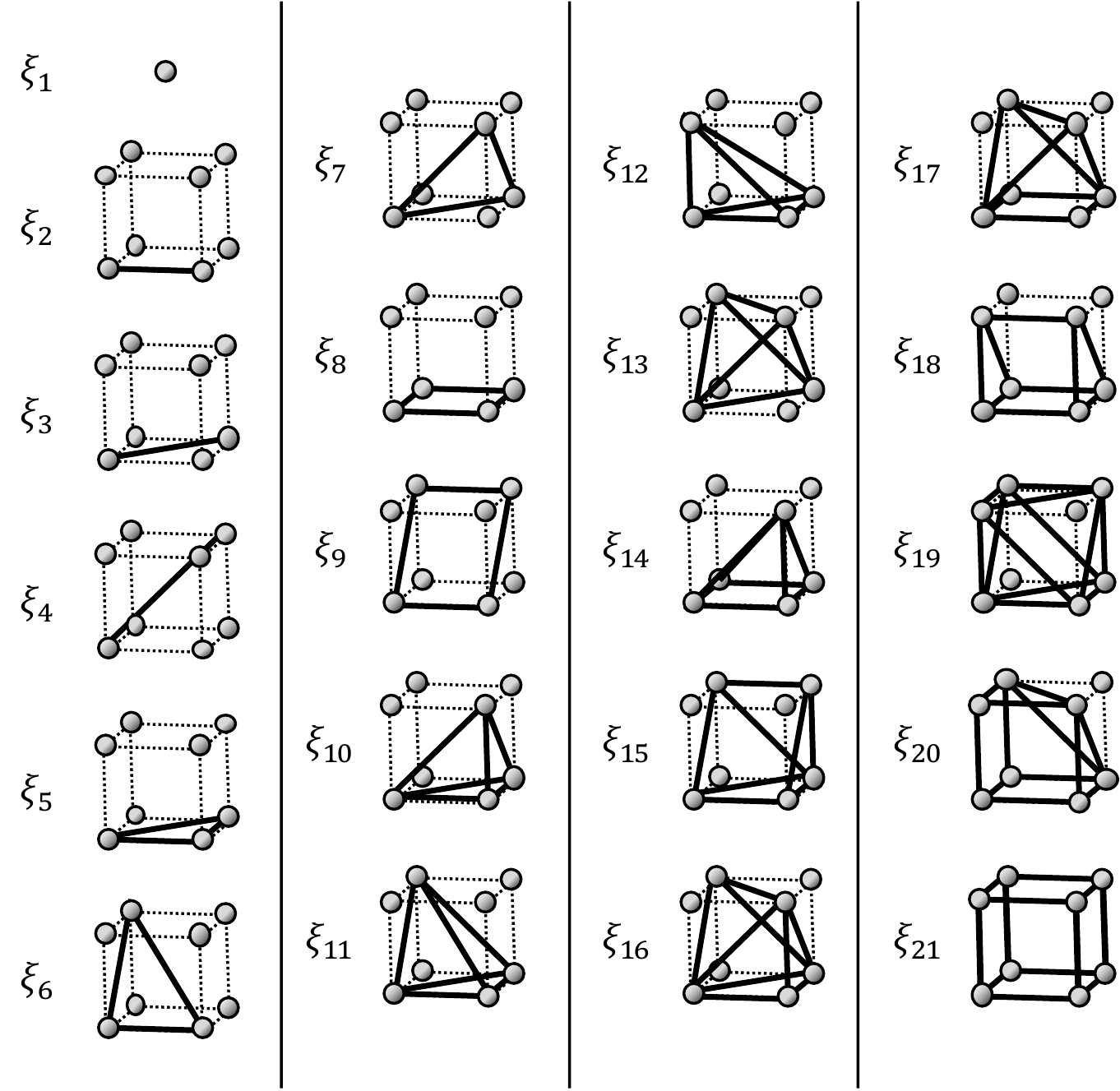}
 \\ \hline \hline
\end{tabular}
}
\end{center}
\end{table}

Note that with the use of the effective cluster interaction energies, $v_m$, and the correlation functions, $\xi_m$, a total energy of a system can be expressed as
\begin{equation}
E= \sum_{m} v_m \xi_m  \; .  \label{eq:CEM_total_energy}
\end{equation}
Here, the superscripts of the sublattices, $\alpha$ and $\beta$, are omitted. Eq.\;\eqref{eq:CEM_total_energy} corresponds to Eq.\;\eqref{eq:total_energy} if the all multi-body effective interaction energies up to the cubic one are taken into account. The correlation functions and the cluster probabilities are directly connected each other as seen in Ref.\;\cite{mohri2013cluster}. One of the biggest advantages of the use of the correlation functions instead of the cluster probabilities is that it can significantly reduce the number of variables used in the calculations. Although we did not encounter any computational issues related to the huge number of variables in this work, for a more demanding calculation (such as the continuous-displacement CVM \cite{kikuchi1998space}, or CDCVM, in the three dimensional lattice), the replacement of the cluster probabilities with the correlation functions would be required.

\end{appendices}

\nocite{apsrev41Control}
\bibliographystyle{apsrev4-2}
\bibliography{ref}

\end{document}